\pdfoutput=1

\documentclass[11pt]{article}
\usepackage{graphicx}
\usepackage{amsfonts}
\usepackage{amssymb,amsmath}
\usepackage{latexsym}
\usepackage{color}
\usepackage{cite}
\usepackage[colorlinks=true,linkcolor=blue,citecolor=blue]{hyperref}
\input{colordvi.tex}

\renewcommand{\thefootnote}{\#\arabic{footnote}}

\newcommand{\bea}{\begin{eqnarray}}   
\newcommand{\eea}{\end{eqnarray}}
\newcommand{\bear}{\begin{array}}  
\newcommand {\eear}{\end{array}}
\newcommand{\bef}{\begin{figure}}  
\newcommand {\eef}{\end{figure}}
\newcommand{\bec}{\begin{center}}  
\newcommand {\eec}{\end{center}}

\begin{document}

\setcounter{footnote}{0}

\renewcommand{\thepage}{\arabic{page}}
\setcounter{page}{1}
\renewcommand{\thefootnote}{\#\arabic{footnote}}

\begin{titlepage}

\setcounter{page}{1} \baselineskip=15.5pt \thispagestyle{empty}

\begin{flushright}
TU-931
\end{flushright}
\vfil

\bigskip\
\begin{center}
{\LARGE \textbf{Spectator field models in light of spectral index \\ after Planck}} \vskip 15pt
\end{center}

\vspace{0.5cm}
\begin{center}
{\Large 
Takeshi Kobayashi,$^{a,b}$\footnote{takeshi@cita.utoronto.ca}
Fuminobu Takahashi,$^{c}$\footnote{fumi@tuhep.phys.tohoku.ac.jp}
Tomo Takahashi,$^{d}$\footnote{tomot@cc.saga-u.ac.jp}\\
and Masahide Yamaguchi$^{e}$\footnote{gucci@phys.titech.ac.jp}
}
\end{center}

\vspace{0.3cm}

\begin{center}
\textit{$^{a}$ Canadian Institute for Theoretical Astrophysics,
 University of Toronto, \\ 60 St. George Street, Toronto, Ontario M5S
 3H8, Canada}\\ 

\vskip 4pt
\textit{$^{b}$ Perimeter Institute for Theoretical Physics, \\ 
 31 Caroline Street North, Waterloo, Ontario N2L 2Y5, Canada}\\ 

\vskip 4pt
\textit{$^{c}$ Department of Physics, Tohoku University, Sendai 980-8578, Japan}

\vskip 4pt
\textit{$^{e}$ Department of Physics, Saga University, Saga 840-8502, Japan}

\vskip 4pt
\textit{$^{e}$ Department of Physics, Tokyo Institute of Technology, Tokyo
152-8551, Japan}
\end{center} \vfil



\begin{abstract}
We revisit spectator field models including curvaton and modulated
reheating scenarios, specifically focusing on their viability in the new
Planck era, based on the derived expression for the spectral index in
general spectator field models.  Importantly, the recent Planck
observations give strong preference to a red-tilted power spectrum,
while the spectator field models tend to predict a scale-invariant one.
This implies that, during inflation,  either (i) the Hubble parameter
varies significantly as in chaotic inflation, or (ii) a scalar potential
for the spectator field has a relatively large negative curvature.
Combined with the tight constraint on the non-Gaussianity, 
the Planck data provides us with rich implications for
various spectator field models.
\end{abstract}

\end{titlepage}

\renewcommand{\thepage}{\arabic{page}}
\setcounter{page}{1}
\renewcommand{\thefootnote}{\#\arabic{footnote}}

\bigskip

\section{Introduction}
\label{sec:1}

The origin of primordial curvature fluctuations is one of the most
important issues in cosmology and has been the target of intense
study. Although  quantum fluctuations of the inflaton are the most
plausible candidate from a minimalist point of view, there may be many other
light scalar fields in nature, one of which is responsible for the
curvature  fluctuations as in the curvaton \cite{Enqvist:2001zp} and
the modulated reheating \cite{Dvali:2003em}  and so on.  
We revisit such spectator field models, specifically focusing on its viability
in the new Planck era. 

The statistical information of the curvature perturbations can be extracted by
evaluating correlation functions.  The power spectrum of the curvature perturbations 
is usually characterized by its amplitude and spectral index,
$\mathcal{P}_\zeta$ and $n_s$,  
which were recently determined by the Planck satellite with unprecedented
accuracy as~\cite{Ade:2013rta},
\bea
\log (10^{10} \mathcal{P}_\zeta) &=& 3.089^{+0.024}_{-0.027}~~ (68 \% \,{\rm CL}) ,\\
n_s &=& 0.9603 \pm 0.0073~~ (68 \% \,{\rm CL}) ,
\label{ns}
\eea
from Planck + WMAP polarization data, 
assuming the standard flat six parameter $\Lambda$CDM model.
Importantly, the Planck data strongly favors a red-tilted power spectrum, i.e., $n_s < 1$, over 
the scale-invariant one at more than $5\sigma$ level.  
On the other hand,
the bispectrum of the curvature perturbations is characterized by non-linearity parameters
for various configurations of the associated three wave numbers. Among them,  it is the non-linearity parameter
of local type, $f_{\rm NL}^{\rm local}$, that is relevant for our purpose.
The Planck data put an extremely stringent constraint on  $f_{\rm NL}^{\rm local}$ as~\cite{Ade:2013tta} 
\bea
f_{\rm NL}^{\rm local} &=& 2.7 \pm 5.8~~ (68 \% \,{\rm CL}). 
\label{Plfnl}
\eea
Our primary concern in this paper is to study the implications of those latest Planck results for the
spectator field models. 

In the spectator field models, there is a light scalar field, $\sigma$, in addition to
the inflaton which drives inflation, such that its quantum fluctuations 
account for the total curvature perturbations. The size of the quantum fluctuations is
determined by the Hubble parameter during inflation, $H_{\rm inf}$. 
If the mass of $\sigma$ is much lighter than $H_{\rm inf}$, 
it hardly evolves during inflation, and therefore, the predicted spectral index is naively 
expected to be close to  unity, since the only source of the scale-dependence is
$H_{\rm inf}$, which however does not evolve significantly during inflation.
Specifically,   the spectral index in the curvaton scenario is given by \cite{Lyth:2001nq}
\begin{equation}
  n_s - 1 = 2 \frac{\dot{H}_{\ast}}{H_{\ast}^2}
                 + \frac23 \frac{V''(\sigma_{\ast})}{H_{\ast}^2}.
  \label{eq:index}
\end{equation}
where $\sigma$ is a curvaton, $V(\sigma)$  the curvaton potential, and
$H$  the Hubble parameter. The prime represents the derivative with
respect to  $\sigma$ and the asterisk indicates that the
variables are evaluated at the horizon exit of the cosmological scales.
One can see from the above formula that the spectral index would be indeed close to unity,
if the curvature of the potential was much smaller than $H_{\ast}$ and if the Hubble
parameter during inflation remained almost constant. 
The same formula also applies to the modulated reheating scenario simply by
identifying $\sigma$ with a modulus that modulates the inflaton decay rate \cite{Kobayashi:2011hp}. 
Compared to the observed red-tilted spectral index (\ref{ns}),  however, it is now clear that such naive estimate is 
insufficient to describe the Planck data, and we need to consider seriously the implications 
of the Planck results  for the spectator field models.\footnote{
Note that the red-tilted spectrum index was already favored by the pre-Planck observations such as WMAP~\cite{Hinshaw:2012fq}.
The implications for the curvaton model were studied in Refs.~\cite{Kawasaki:2011pd,Kawasaki:2012gg}.
} 

To this end, we first present a simple derivation showing that the above
formula (\ref{eq:index}) is rather robust and holds in a general
spectator field  model, 
as long as the light scalar field is responsible for the total curvature
perturbations and it does not affect the dynamics of inflation at least
until the CMB scales exited the horizon. 
The spectral index for some other cases are discussed as well.
This enables us to derive implications  from the Planck result on $n_s$
for various spectator field models such as those studied in
Ref.~\cite{Suyama:2010uj}.
We also refer the reader to~\cite{Byrnes:2006fr} which derived the
expression~(\ref{eq:index}) under general situations. 

The predicted  expression for $n_s$ in a general
spectator field model  implies that
there are basically two choices to realize the 
observed red-tilted spectral index. The first one is to consider a relatively
large variation of the Hubble parameter during inflation.  For example,
a chaotic inflation based on a quartic potential is able to account for $n_s \simeq 0.96$. 
The second choice is to assume a
relatively large and negative mass squared of the potential. Specifically, a potential with 
a negative curvature satisfying $V''(\sigma_{\ast}) \simeq - 0.06 H_*^2$ leads to $n_s \simeq 0.96$. 
As we shall discuss in detail in Sec.~\ref{sec:3} and  Sec.~\ref{sec:curvaton}, 
this places non-trivial constraints on the models. In the curvaton scenario, for instance,
the curvaton should come close to dominating the Universe when it decays, in order to satisfy the
stringent constraint on non-Gaussianity (\ref{Plfnl}). However this turns out to be quite non-trivial, because
of the large negative curvature; the curvaton tends to start to oscillate relatively soon after inflation ends,
and it is considered to decay faster for a heavier curvaton mass.
Then the curvaton domination may be realized only for high reheating temperature and high-scale inflation.
This problem can be circumvented by invoking a hilltop curvaton model~\cite{Kawasaki:2008mc,Kawasaki:2011pd,Kawasaki:2012gg},
where the onset of the oscillation is delayed. The curvaton
domination is then allowed with many orders of magnitude of the inflation and reheating scales. 
Interestingly, however, $f_{\rm NL}^{\rm local}$ is generically predicted to be of ${\cal O}(10)$ in the hilltop 
limit~\cite{Kawasaki:2011pd}, which will be in tension with (\ref{Plfnl}). The tension can be ameliorated
if the curvaton drives a short duration of inflation before it starts to oscillate~\cite{Kawasaki:2012gg}. Such argument already shows the rich implications
of the Planck results.  We will discuss these issues in depth in Sec.~\ref{sec:3} and  Sec.~\ref{sec:curvaton}.

Let us mention briefly the implications of (\ref{Plfnl}).
It has been often claimed that large non-Gaussianities of the curvature
perturbations can be produced in spectator field models such as the curvaton
\cite{Lyth:2002my,Lyth:2003ip,Bartolo:2003jx,Sasaki:2006kq} and the
modulated reheating models~\cite{Zaldarriaga:2003my,Suyama:2007bg}. 
However, in fact, the spectator field models predict a $f_{\rm NL}^{\rm local}$
of order unity for a wide (therefore natural) parameter space~\cite{Suyama:2013nva},
and so,  the constraint on the non-Gaussianity alone does not exclude
the spectator field models. As we have seen above, it is very important to
study the implications of both (\ref{ns}) {\it and} (\ref{Plfnl}) for the spectator field models. 
The prediction for $f_{\rm NL}^{\rm local}$  of order unity should be contrasted to
that of the standard single field inflation models in which an
inflaton itself is responsible for the curvature perturbations,  because
the latter predicts $f_{\rm NL}^{\rm local}$ suppressed by  slow-roll parameters~\cite{Maldacena:2002vr,Creminelli:2004yq}.
There is still room for non-zero primordial non-Gaussianity satisfying (\ref{Plfnl}).

Our paper is organized as follows: In the next section, we derive a
 formula of the spectral index that holds in a  general spectator field model. In
Sec.~\ref{sec:3}, we discuss how to accommodate the spectator field models 
to the Planck results, particularly paying attention to the spectral index of the
curvature perturbations. In Sec.~\ref{sec:curvaton}, we discuss the implications for
the curvaton scenario in detail.
The last section is devoted to discussion and conclusions.

\section{Spectral index in spectator field models}
\label{sec:2}

In this section we derive a general expression for the spectral index of
the curvature perturbation spectrum produced by spectator fields.
Here we define spectator field models as mechanisms where the curvature
perturbations are generated by a light scalar~$\sigma$ having no
influence on the inflationary expansion while the CMB scales exit the
horizon. (Hence we do not consider fluctuations of the inflaton
field. In other words, we assume that the inflaton-induced perturbations
are subdominant compared to those from the spectator fields.)
Examples of such models include the curvaton~\cite{Enqvist:2001zp}, 
modulated reheating~\cite{Dvali:2003em}, and inhomogeneous end of
inflation~\cite{Bernardeau:2002jf,Bernardeau:2004zz,Lyth:2005qk,Salem:2005nd,Alabidi:2006wa,Kawasaki:2009hp}.

Using the $\delta \mathcal{N}$-formalism
\cite{229378,astro-ph/9507001,astro-ph/0003278,astro-ph/0411220}, the
curvature perturbation~$\zeta$ can be computed as the difference among
different patches of the Universe in the e-folding numbers between an
initial flat time-slice when the separate universe assumption is a good
approximation, and a final uniform-density time-slice when the Universe
is adiabatic:
\begin{equation}
 \zeta_{\boldsymbol{k}} = \frac{\partial \mathcal{N}}{\partial \sigma }
  \delta \sigma_{\boldsymbol{k}},
\label{deltaN}
\end{equation}
where we have expanded~$\zeta$ in Fourier space in terms of the field
fluctuations to linear order, and the background is assumed to be
homogeneous and isotropic.  Here we consider the Universe to be an
attractor-like system, so that the e-folding number is given as a function of
the field value~$\sigma$ at an arbitrary time when $\sigma$ follows an
attractor solution, i.e., $ \mathcal{N} =
\mathcal{N} (\sigma) $.

Now, let us choose two instants of time: $t_*$ when a certain wave
number~$k$ (say, the pivot scale) exits the horizon (i.e. $k = aH$), and
$t_0$ when the smallest 
among the wavelengths of interest (say, the entire CMB scales) exits the
horizon. Then the curvature perturbation can be written as
\begin{equation}
 \zeta_{\boldsymbol{k}} = \frac{\partial \mathcal{N}_*}{\partial
  \sigma_0} \frac{\partial \sigma_0}{\partial \sigma_*} \delta
  \sigma_{\boldsymbol{k}* },
\label{eq3}
\end{equation}
where $\mathcal{N}_*$ is the e-folding number between time~$t_*$
and the final time-slice at~$t_f$,
\begin{equation}
 \mathcal{N}_* = \int^{t_f}_{t_*} H dt,
\end{equation}
and for the other quantities the subscripts~$*$ and $0$ denote values at
$t_*$ and $t_0$, respectively.
Since we have assumed that the field~$\sigma$ has negligible effects on
the inflationary expansion while the wave modes of interest exit the
horizon, (\ref{eq3}) can be rewritten using
\begin{equation}
 \mathcal{N}_0 = \int^{t_f}_{t_0} H dt
\end{equation}
as
\begin{equation}
 \zeta_{\boldsymbol{k}} = \frac{\partial \mathcal{N}_0}{\partial
  \sigma_0} \frac{\partial \sigma_0}{\partial \sigma_*} \delta
  \sigma_{\boldsymbol{k}* }.
\end{equation}
Note that this expression depends on the wave number~$k$ only through
$\partial \sigma_0 
/ \partial \sigma_*$ and $\delta \sigma_{\boldsymbol{k}*}$. 

Defining the power spectrum~$\mathcal{P}_\zeta (k)$ of the curvature
perturbations as
\begin{equation}
 \langle \zeta_{\boldsymbol{k}} \zeta_{\boldsymbol{k'}} \rangle
 = 
 (2 \pi)^3 \delta^{(3)} (\boldsymbol{k} + \boldsymbol{k'}) 
 \frac{2 \pi^2}{k^3} \mathcal{P} (k),
\end{equation}
then it can be expressed in terms of the power spectrum of the field
fluctuations~$\mathcal{P}_{\delta \sigma}(k)$ (defined similarly as
above) as
\begin{equation}
 \mathcal{P}_\zeta (k) = 
\left(
\frac{\partial \mathcal{N}_0}{\partial
  \sigma_0} \frac{\partial \sigma_0}{\partial \sigma_*}\right)^2 
 \mathcal{P}_{\delta \sigma *}(k).
\label{P8}
\end{equation}

We suppose that during inflation the Hubble parameter is nearly
constant, and that the light field~$\sigma$ follows
the slow-roll attractor under a potential~$V(\sigma)$ which is a
function merely of~$\sigma$, 
\begin{equation}
 3 H \dot{\sigma} \simeq - V'(\sigma ),
\label{slow-roll}
\end{equation}
where an overdot represents a time-derivative, and a prime a
$\sigma$-derivative.
Here we are assuming that the classical rolling of~$\sigma$ dominates
over the quantum fluctuations,\footnote{For cases where the quantum
fluctuations dominate, a stochastic approach may be required. See
e.g.~\cite{Finelli:2010sh}.} 
\begin{equation}
 \frac{|\dot{\sigma} |}{H}  \simeq \frac{|V'(\sigma)|}{3 H^2} > \frac{H}{2 \pi }.
\end{equation}
Then by integrating (\ref{slow-roll}), one obtains 
\begin{equation}
 \int^{\sigma_0}_{\sigma_*} \frac{d \sigma}{V'(\sigma) } = - \int^{t_0}_{t_*}
  \frac{dt}{3 H}.
\label{P13}
\end{equation}
Here we stress again that the right hand side is independent
of~$\sigma$, hence differentiating both sides in terms of~$\sigma_*$
gives
\begin{equation}
 \frac{\partial \sigma_0}{\partial \sigma_*} =
  \frac{V'(\sigma_0)}{V'(\sigma_*)}.
\end{equation}
Further using
\begin{equation}
 \mathcal{P}_{\delta \sigma *} (k) = \left( \frac{H_*}{2 \pi} \right)^2,
\end{equation}
then (\ref{P8}) becomes
\begin{equation}
 \mathcal{P}_{\zeta} (k) = \left( \frac{\partial \mathcal{N}_0}{\partial
			    \sigma_0} 
  \frac{V'(\sigma_0)}{V'(\sigma_*)} \right)^2
 \left(\frac{H_*}{2\pi}\right)^2,
\end{equation}
where now its scale-dependence shows up through $V'(\sigma_*)$ and
$H_*$. 
Thus, using $ d \ln k \simeq H_* dt$ and (\ref{slow-roll}), one
arrives at a general expression for the spectral index in
spectator field models:
\begin{equation}
 n_s - 1 \equiv \frac{d \ln \mathcal{P}_{\zeta}}{d \ln k} = 
 2 \frac{\dot{H}_*}{H_*^2} + \frac{2}{3} \frac{V''(\sigma_*)}{H_*^2} . 
\label{spectral_index}
\end{equation}
This clearly shows that a red-tilted (i.e. $n_s < 1$) spectrum is
explained either by a tachyonic potential $V'' < 0$, or 
a time-varying Hubble parameter $\dot{H} < 0$ during inflation.
The expression~(\ref{spectral_index}) has also been derived
in~\cite{Byrnes:2006fr} through a generic treatment of
adiabatic/isocurvature fluctuations. 
See also~\cite{Sasaki:1995aw}. We further note that~(\ref{spectral_index})
can be obtained from the generic expressions for isocurvature
perturbations derived in~\cite{Polarski:1994rz}.

\vspace{\baselineskip}

The above result can also be extended to cases with multiple light
fields~$\sigma^a$ ($a = 1,\, 2,\, \cdots$), where each field has no effect on the
inflationary expansion while the scales of interest exit the horizon. 
Further assuming that the potential of each field is decoupled from the
others, i.e. $V = \sum_{a} V_a (\sigma^a)$, and also that 
the fluctuations between different fields have no correlations, 
one can check that
\begin{equation}
 \mathcal{P}_{\zeta} (k) = \sum_{a}
\left( \frac{\partial \mathcal{N}_0}{\partial
			    \sigma^a_0} 
  \frac{V_a'(\sigma_0^a)}{V_a'(\sigma^a_*)} \right)^2
 \left(\frac{H_*}{2\pi}\right)^2,
\end{equation}
where now a prime on $V_a(\sigma^a)$ denotes differentiation in terms
of~$\sigma^a$. Hence introducing 
\begin{equation}
 q^a \equiv \left(
\frac{\partial \mathcal{N}_0}{\partial \sigma_0^a}
\frac{V_a'(\sigma_0^a)}{V_a'(\sigma_*^a)}
\right)^2
\bigg/
\sum_{b}
\left(
\frac{\partial \mathcal{N}_0}{\partial \sigma_0^b}
\frac{V_b'(\sigma_0^b)}{V_b'(\sigma_*^b)}
\right)^2,
\end{equation}
(which satisfies $\sum_{a} q^a = 1$), one obtains the expression
\begin{equation}
  n_s - 1 =  2 \frac{\dot{H}_*}{H_*^2} + 
  \frac{2}{3}  \sum_{a} q^a  \frac{ V_a''(\sigma^a_*)}{H_*^2} 
\end{equation}
for multiple spectator fields.

\vspace{\baselineskip}

Similar analyses can be carried out also for 
potentials~$V(\sigma, t)$ that explicitly depend on time, especially 
when it takes a separable form~\cite{Kobayashi:2013nva} during inflation,\footnote{The
potential can further have time-independent terms that are negligible
during inflation, but play important roles in the post-inflationary epoch.} i.e.,
\begin{equation}
 V(\sigma, t) =  v(\sigma) f(t) .
\end{equation}
(A good example is a Hubble-induced mass term $V \propto  H^2
\sigma^2$.)
Then given that the time-dependent part varies slowly 
\begin{equation}
  \left| \frac{\dot{f}}{H f}   \right| \ll 1,
\end{equation}
the light field  (i.e. $| v'' f / H^2 | \ll 1$) follows the slow-roll
approximation~(\ref{slow-roll}) and now (\ref{P13}) can be rewritten as
\begin{equation}
  \int^{\sigma_0}_{\sigma_*} \frac{d \sigma}{v'(\sigma) } = - \int^{t_0}_{t_*}
  \frac{f \, dt}{3 H}.
\label{P23}
\end{equation}
If $f(t)$ is unaffected by~$\sigma$ (for e.g. when $f(t)$ arises from
a Hubble-dependence), then so is the right hand side of (\ref{P23}), and thus
one reproduces the same expression as (\ref{spectral_index}):
\begin{equation}
 n_s - 1  =  2 \frac{\dot{H}_*}{H_*^2} + 
 \frac{2}{3} \frac{v''(\sigma_*) f(t_*)}{H_*^2} . 
\end{equation}

\section{Red-tilted spectrum in spectator field models}
\label{sec:3}

In this section, we discuss how we can construct spectator field models such
as the curvaton and the modulated reheating scenarios consistent with
the Planck data, in particular, from the view point of the spectral
index $n_s$.  As mentioned in the introduction, models with a light
spectator field tend to give a rather scale-invariant spectral index, which
is now significantly away from the Planck value. In this section, we discuss two
possibilities to directly explain the spectral index $n_s = 0.9603 \pm
0.0073$ by the Planck results in the context of the spectator field models:
(A) Relatively large variation of the Hubble parameter (B) Negative mass
squared of the potential.

\subsection{Variation of the Hubble parameter}

If the first term of the right hand side in Eq.~(\ref{spectral_index}), that is, if $2 \dot{H}_{\ast}/H_{\ast}^2$ is
non-negligible, the spectral index can deviate from unity. 
When inflation is driven by a canonical slow-rolling inflaton, 
one can rewrite as
\begin{equation}
   2 \frac{\dot{H}_{\ast}}{H_{\ast}^2} = -2\epsilon
    = - \frac{1}{M_p^2} \left( \frac{d\phi}{dN} \right)^2,
   \label{eq:2epsilon}
\end{equation}
where $\phi$ is the inflaton, $N$ is the e-folding number, $\epsilon$ is
the slow-roll parameter, and $M_p \simeq 2.4 \times 10^{18}$~GeV
is the reduced Planck mass. Thus, if this term is responsible for $1-n_s
\simeq 0.04$, $\Delta\phi \simeq \sqrt{1-n_s} N M_p > M_p$ for $N \sim 50$, which requires large field inflation.  Actually,
in case of chaotic inflation with the potential $V(\phi) \propto
\phi^p$, $2\epsilon \simeq p^2 M_p^2 /\phi^2 \simeq p/(2N)$. Thus, $n_s \simeq
0.96$ implies $p \sim 2 (1-n_s) N \sim 4$ for $N \sim 50$. Thus,
the chaotic inflation with a quartic potential can nicely explain the observed
red-tilted spectral index.\footnote{
It is interesting
to notice that such chaotic inflation based on a quartic potential
naturally appears by the use of D-term in supergravity \cite{Dterm}, though
it is realized by F-term as well\cite{Fterm}.
} 
This should be contrasted to the fact that the chaotic inflation with a quartic potential
 is already strongly disfavored in the case where the inflaton is responsible for the density perturbations because
of too large tensor-to-scalar ratio $r$. In spectator field models,  however, 
the tensor-to-scalar ratio is often suppressed (or equivalently, the curvature perturbations are enhanced),
making such inflation models 
consistent with the observations.\footnote{
Constraints on inflation models in spectator field models 
have been investigated in \cite{mixed}.
}
\footnote{
It has recently been pointed out in~\cite{Kobayashi:2013awa}
that in some cases, spectator fields can also suppress the inflaton-induced curvature
perturbations and thus allow the tensor-to-scalar ratio~$r$ to be much larger
than $(d\phi / dN)^2 / M_p^2$.
}

Finally, let us comment on k-inflation with ${\cal L} = K(\phi,X)$
\cite{ArmendarizPicon:1999rj}. The far right hand side in
Eq. (\ref{eq:2epsilon}) is multiplied by $K_{,X}$, that is, $-2\epsilon =
- (K_{,X}/M_p^2) (d\phi/dN)^2$. Here $X = (d\phi/dt)^2/2$ and
$K_{,X}$ is the derivative with respect to $X$. Therefore, if $|K_{,X}|
\gg 1$, large field inflation may not necessarily be required. However,
for example, in case of the DBI inflation \cite{Silverstein:2003hf},
$K_{,X} = c_s^{-1}$ and the sound velocity $c_{s}$ is now strongly
constrained from the non-Gaussianity as $c_s^{\rm DBI} \ge 0.07$ (95\% CL)
\cite{Ade:2013tta}. Therefore, large variation comparable to the Planck
scale is still required, which is difficult to realize from the
microscopic view point \cite{Baumann:2006cd,Lidsey:2007gq,Kobayashi:2007hm}.

\subsection{Negative mass squared}
\label{subsec:Nms}

Let us next consider the case where the second term in the right hand side of
Eq.~(\ref{spectral_index}), i.e.  $2V''(\sigma_{\ast})/(3H_{\ast}^2)$, is responsible for
the deviation of the spectral index from the scale invariant one.
 In this case, the observed  red-tilted spectrum implies a negative mass squared of the
effective scalar potential, $V''(\sigma_*) < 0$. 
More concretely, $n_s \simeq 0.96$ can be realized if 
$V''(\sigma_{\ast}) \simeq - 0.06 H_{\ast}^2$.

Here, we discuss implications of the negative mass squared for the spectator field models.
For the moment we assume that the scalar potential $V(\sigma)$ does not depend 
on time. 
The known spectator field models can be broadly classified into two cases, depending on
whether  the spectator field fluctuations are converted into curvature perturbations
before or after the commencement of oscillations of~$\sigma$. 
The former  includes 
the modulated reheating~\cite{Dvali:2003em}, the inhomogeneous end of
inflation~\cite{Bernardeau:2002jf,Bernardeau:2004zz,Lyth:2005qk,Salem:2005nd,Alabidi:2006wa,Kawasaki:2009hp},
the modulated trapping~\cite{Langlois:2009jp}, velocity modulation\cite{Nakayama:2011bc}, and so on (i.e. almost all the
spectator field models except the curvaton model), while the latter contains the curvaton mechanism as well as 
some realization of the modulated reheating. 

In the former class of models, the spectator field should not start to oscillate until a certain point, as
the resultant curvature perturbations would be strongly suppressed, otherwise. 
Suppose that it starts to oscillate when the Hubble parameter becomes comparable to 
 the absolute magnitude of the effective mass, $H^2 \sim |V''|$. Then, the required relation 
 $V''(\sigma_{\ast}) \simeq - 0.06 H_{\ast}^2$ implies that the spectator field $\sigma$ starts to oscillate relatively soon after inflation. 
This does not place any stringent constraints on the scenario of inhomogeneous end of inflation.
 On the other hand, however, 
the modulated reheating scenario is severely constrained,
because the reheating should be completed soon after inflation, before the modulus
starts to oscillate. 
This implies that the reheating process must be extremely efficient, which
results in rather high reheating temperature
for most of inflation models. Further, it is unclear whether it is possible to induce such an efficient reheating process
via a usual perturbative decay of the inflaton, on which most of the calculations of the modulated reheating 
are based.

It should be noticed that the above discussion is based on the assumptions
that the scalar potential of the spectator field is time-independent and
it starts to oscillate when $H^2 \sim |V''|$. If these assumptions are relaxed,
the commencement of oscillations can be delayed, avoiding the above constraint.
 First,  consider a time-dependent scalar potential.
For example,  the spectator field may receive a negative Hubble mass correction,
$- (k/2) H^2 \sigma^2$, from  interactions with the inflaton, where $k>0$ is a numerical coefficient.
Indeed, such corrections arise ubiquitously in supergravity. If $k$ is about $0.06$,
the observed spectral index can be explained.  Importantly, the Hubble mass
decreases after inflation. After reheating, the inflaton contribution to the Hubble mass disappears, 
but instead, a Hubble induced mass with $k={\cal O}(10^{-2})$ generically arises from Planck-suppressed couplings
with the standard-model particles that constitute the thermal plasma~\cite{Kawasaki:2011zi}.
Therefore, the early oscillations of the spectator field can be avoided, if the zero-temperature
potential has a  curvature much smaller than the Hubble parameter during inflation. 
Next let us consider a hilltop initial condition for the spectator field.
As long as the spectator field sits sufficiently near the top of the potential~\footnote{
If it sits too close to the top of the potential, it may lead to the formation of topological defects.
 },  the onset of oscillations will be delayed. This
can be understood as follows. The spectator field
starts to oscillate when
\begin{equation}
  \left|\frac{\dot{\sigma}_{\rm osc} }{ H_{\rm osc} (\sigma_{\rm osc} - \sigma_{\rm min}) } \right| \sim 1,
\end{equation}
where the potential minimum is located at $\sigma = \sigma_{\rm min}$, and we assume that
the potential monotonically increases from the potential minimum to the maximum.
Under the slow-roll approximation, the above relation can be rewritten as
\begin{equation}
  H_{\rm osc}^2 \sim \left| \frac{V'(\sigma_{\rm osc}) }{ (\sigma_{\rm osc} - \sigma_{\rm min})}\right|.
\end{equation} 
 Therefore, the onset of the oscillation of a spectator field $\sigma$ is
significantly delayed compared to the naive expectation, if
\begin{equation}
  |V'(\sigma_{\rm osc})| \ll | (\sigma_{\rm osc} - \sigma_{\rm min})
   V''(\sigma_{\rm osc})|.
\label{Gu28}
\end{equation} 
This condition is satisfied if $\sigma$ sits near the top of the potential.
Thus, the hilltop initial condition relaxes the constraint on the spectator models.\footnote{ In principle it is possible 
to combine the two possibilities mentioned
 (i.e., negative Hubble-induced mass + hilltop initial condition). }

The representative example of the latter class of the spectator field
models, i.e. cases where the conversion of the field fluctuations into
curvature perturbations happens after the onset of field oscillations, 
is the curvaton model~\cite{Enqvist:2001zp}. 
The curvature perturbations are generated  effectively when the
curvaton starts to (or at least comes close to) dominate the Universe, and then 
 the curvaton decay produces a significant amount of entropy and radiation. Importantly,
 it is the density perturbations, $\delta \rho_\sigma/\rho_\sigma$, that is relevant for
evaluating the resultant curvature perturbations. Note that both $\delta \rho_\sigma$ and 
$\rho_\sigma$ decrease after the onset of oscillations.
 This is the reason why the curvaton
model works well even though the conversion takes place after the onset of
the curvaton oscillation. In order to estimate the final curvature perturbations, however, it is 
necessary  to follow the evolution of the curvaton from during inflation until its decay, which 
requires a dedicated analysis in case of the negative mass squared. Therefore, we
devote the next section to the detailed analysis of this case. 

Finally let us briefly mention another example in the latter class of the spectator field models.
Usually it is assumed that the modulus does not start to oscillate until reheating in the modulated
reheating scenario. However, some realization of the model may work even if the 
modulus starts to oscillate before the reheating. This is the case if the inflaton decay rate is
proportional to some power of $\sigma$. For instance, suppose that the inflaton decays through an interaction whose coefficient
 is proportional to $\sigma$, in which case the inflaton decay rate is proportional to $\sigma^2$.
As long as the decay mode gives dominant contribution to the total decay rate,
and given that the fluctuation of the decay rate is determined by the oscillation
amplitude, the resultant curvature perturbation 
$\zeta \sim \delta \Gamma /\Gamma \propto \delta \ln \sigma_{\mathrm{amp}}^2$
is not significantly suppressed even after the commencement of oscillations. 
In this case, we need to follow the
evolution of the modulus during from inflation until reheating in order to estimate the final curvature perturbation.
To this end, the technique used in the curvaton scenario, which we shall discuss in the next section, will be useful,
and in fact it can be applied to this class of the modulated reheating in a straightforward way. 
Note that the problems associated with the early oscillations can be avoided in this case, allowing lower reheating
temperature.

\section{Planck constraints on curvaton scenarios}
\label{sec:curvaton}

In the curvaton scenario~\cite{Enqvist:2001zp}, the spectator
field~$\sigma$ (which we refer to as the curvaton in this section)
undergoes oscillations around its potential minimum in the
post-inflationary era. 
After the inflaton decays, the curvaton's energy density increases
relative to the background radiation, and thus the curvaton generates curvature
perturbations as it comes close to dominating the Universe. 
The curvaton is assumed to eventually decay into radiation. 

As is seen in (\ref{spectral_index}),
the curvaton needs to lie along a
negatively curved potential during inflation 
with an effective mass not much smaller than the inflationary Hubble parameter,
\begin{equation}
 V''(\sigma_*) \sim - 10^{-2 } H_*^2,
\label{curvaton_mass}
\end{equation}
in order to produce the red-tilt (\ref{ns}), if not from large-field inflation. 
However, unlike other spectator mechanisms the curvaton generates
perturbations as it oscillates, thus issues discussed in the previous
section are not necessarily problems here. 
In this section we will see instead that the Planck constraints on 
non-Gaussianity impose rather strict requirements for curvaton
model building.

\subsection{Quadratic curvatons}
\label{subsec:quad_curv}

Let us start by discussing the simplest case, namely curvatons with a
quadratic potential,
\begin{equation}
 V(\sigma) = \frac{1}{2} m_\sigma^2 \sigma^2 . 
\label{quad-pot}
\end{equation}
Although a quadratic curvaton itself cannot produce a red-tilt, this
simple model captures some basic properties that are shared by a rather
wide group of curvaton potentials. 
(For e.g., curvatons with cosine-type potentials~(\ref{axionic_curvaton})
behave similarly to quadratic ones except for when $\sigma_*$
is very close to the potential maximum.)
Hence the discussions in this subsection also apply to curvatons
whose overall potential forms do not drastically deviate from a
quadratic one. 

Denoting the energy density ratio between the curvaton and radiation
(originating from the inflaton) upon curvaton decay by
\begin{equation}
 \hat{r} \equiv \left. \frac{\rho_\sigma}{\rho_r}
		\right|_{\sigma\mathrm{-decay}},
\label{hatr}
\end{equation}
the power spectrum and non-Gaussianity parameter from a quadratic
curvaton takes the form
\begin{equation}
 \mathcal{P}_\zeta = \left( \frac{2 \hat{r}}{4 + 3 \hat{r}} \frac{H_*}{2
		      \pi \sigma_*}  \right)^2,
\label{Pzetaquad}
\end{equation}
\begin{equation}
 f_{\mathrm{NL}}  = \frac{5}{12} 
 \left( -3 + \frac{4}{\hat{r}} + \frac{8}{4 + 3 \hat{r}}   \right).
\label{fNLquad}
\end{equation}
The Planck constraint on $f_{\mathrm{NL}}$~(\ref{Plfnl})
requires the curvaton to (almost) dominate the Universe before decaying,
i.e.,
\begin{equation}
 \hat{r} \gtrsim 0.1.
\label{hatrone}
\end{equation}
A quadratic curvaton starts to oscillate when the Hubble parameter
becomes comparable to its mass, i.e. $H_\mathrm{osc} \sim
m_\sigma$. (Hereafter we use the subscript ``osc'' to denote quantities
at the onset of the curvaton oscillation). Then, ignoring the time
evolution of the curvaton field and 
using $\sigma_\mathrm{osc} \sim \sigma_*$, the curvaton energy fraction
at $t = t_{\mathrm{osc}}$ is
\begin{equation}
 \left. \frac{\rho_\sigma }{\rho_{\mathrm{total}}}
 \right|_{\mathrm{osc}} 
 \simeq \frac{V(\sigma_\mathrm{osc})}{3 M_p^2 H_{\mathrm{osc}}^2 }
 \sim (10^3 \mathrm{-} 10^6) \left(\frac{H_*}{M_p}\right)^2
 \lesssim 10^{-3},
\label{fraction_osc}
\end{equation}
where we have also used (\ref{Pzetaquad}), (\ref{hatrone}), and
$\mathcal{P}_\zeta \approx 2.2 \times 10^{-9}$. The far right hand side
is due to the upper bound  
on the inflationary scale from the constraint on the tensor-to-scalar
ratio $r < 0.11$ (95\% CL)~\cite{Ade:2013rta}. 
One sees from (\ref{fraction_osc}) that the energy fraction at the onset
of the oscillations is bounded by~$H_*$, in other words, inflation with
lower scales requires the curvaton to oscillate for larger numbers of
e-foldings before dominating the Universe. 
Therefore the curvaton scenario can be successful provided that the
inflation and reheating (= inflaton decay) scales are high and/or the
curvaton decay rate is low.

Here it should be noted that the curvaton decay rate typically scales as
some positive power of the oscillation mass. Unless the oscillation mass
is designed to be significantly smaller than the curvaton's effective
mass during inflation~(\ref{curvaton_mass}), the observed value of the spectral
index~(\ref{ns}) requires a rather large curvaton mass and thus 
prevents the curvaton decay rate from being tiny. 
Whether this issue becomes a serious problem depends on the details of the
model, however this was demonstrated in~\cite{Kawasaki:2011pd} to impose
severe constraints when the curvaton is a pseudo-Nambu-Goldstone
boson of a broken U(1) symmetry, possessing a cosine-type potential
\begin{equation}
 V(\sigma) = \Lambda^4 \left[ 1 - \cos \left(\frac{ \sigma }{f} \right)
		       \right].
\label{axionic_curvaton}
\end{equation}
Interactions of such an axionic curvaton with other particles are
typically suppressed by
the symmetry breaking scale~$f$, and thus in terms of the effective mass
at the minimum $m_\sigma = \Lambda^2 / f$, its decay rate reads
\begin{equation}
 \Gamma_\sigma \sim \frac{m_\sigma^3}{f^2} = \frac{\Lambda^6}{f^5}.
\end{equation}
For a large effective mass of (\ref{curvaton_mass}), the axionic
curvaton can dominate the Universe and produce the primordial
perturbations only if the inflationary scale and reheating (= inflaton
decay) temperature
are as high as 
$H_{\mathrm{inf}}$, $T_{\mathrm{reh}} \gtrsim 10^{13} \mathrm{GeV}$
(almost saturating the current bound on primordial tensor modes), or if
the initial curvaton field value is located very close to the
hilltop $\sigma_* \approx\pi f$ such that the system is no longer 
approximated by a quadratic potential. 
This latter case is discussed later.

In summary, high inflation and reheating scales and/or suppression of
the curvaton decay rate are required 
in order for curvatons following the familiar relations
\begin{equation}
 \mathcal{P}_\zeta \sim \left( \frac{H_*}{\sigma_*} \right)^2, \qquad
 H_{\mathrm{osc}} \sim m_\sigma ,
\label{quad-ish}
\end{equation}
to dominate the Universe and produce perturbations consistent with
observations.

\subsection{Possible curvaton scenarios with negative mass squared}
\label{subsec:possible}

Facing the situation noted above, we lay out possible ways to
reconcile the curvaton mechanism with observations without relying on
the inflaton sector. The basic ideas are to extend the curvaton
lifetime and/or enhance the curvaton energy fraction prior to the
oscillations. 

\begin{itemize}
 \item {\it Suppressing the decay rate.} 
The curvaton can
dominate the Universe and produce nearly Gaussian perturbations if its
decay rate is highly suppressed, for e.g., if the decay rate vanishes at
tree level, or is helicity suppressed. Furthermore, a curvaton potential
whose effective mass around its minimum is much smaller than that
at~$\sigma_*$ can suppress the decay rate. 

 \item {\it Potential with a flat plateau.} 
If the negatively curved region of the potential around~$\sigma_*$
during inflation and
the potential minimum is connected by a flat plateau, then the curvaton
would undergo a period of slow-roll and delay the onset of the
oscillations. This can enhance the curvaton energy fraction.

 \item {\it Negative Hubble-induced mass squared.}
Curvatons that obtain negative Hubble-induced mass terms during inflation can
produce a red-tilted perturbation spectrum. Furthermore, the
Hubble-induced mass decreases/vanishes in the post-inflationary era and
thus the onset of the curvaton oscillations is delayed. 

 \item {\it Hilltop curvaton.}
The onset of oscillations is delayed for a curvaton that lies close to a
local maximum of its potential
(cf. Sec.~\ref{subsec:Nms}). Furthermore, hilltop potentials
significantly enhance the linear perturbations~\cite{Kawasaki:2011pd},
and thus violate both relations 
in~(\ref{quad-ish}). However, the latest $f_{\mathrm{NL}}$ bound
from Planck imposes rather strict constraints on such hilltop curvaton
scenarios. Let us elaborate on this case in the next subsection.

\end{itemize}

\subsection{Hilltop Curvatons}

A simple way to resolve the issues discussed in
Sec.~\ref{subsec:quad_curv} and give a tachyonic curvaton
mass~(\ref{curvaton_mass}) is to have the curvaton
located close to a local maximum of its potential during inflation,
i.e., the potential well approximated by
\begin{equation}
 V(\sigma) = V_0 - \frac{1}{2} m_\sigma^2 (\sigma - \sigma_1)^2
\label{hilltopV}
\end{equation}
until the curvaton starts to oscillate
(for e.g. the $\sigma_* \to \pi f$ limit of the axionic curvaton). 
The hilltop potential produces a red-tilted perturbation spectrum, and
since the potential flattens out in the hilltop limit the onset of the
curvaton oscillation is delayed to $H_{\mathrm{osc}}^2 \ll m_\sigma^2 $. 
Furthermore it is known that such flattened potentials lead to an
inhomogeneous onset of the curvaton oscillation, and as a consequence
strongly enhance the resulting curvature
perturbations~\cite{Kawasaki:2011pd,Kawasaki:2012gg}.
Thus hilltop curvatons simultaneously violate the relations in
(\ref{quad-ish}) and can dominate the Universe without requiring
high scale inflation or suppression of~$\Gamma_\sigma$.\footnote{On the
other hand, steep potentials such as in self-interacting
curvatons~\cite{Enqvist:2005pg,Enqvist:2008gk,Enqvist:2009ww,Kobayashi:2012ba}
tend to suppress the resulting
perturbations compared to quadratic cases. This further suppresses the
energy fraction upon decay~(\ref{fraction_osc}).}

Let us set the potential minimum (existing outside the field range
where (\ref{hilltopV}) is a good approximation) to $\sigma = 0$, and
without loss of generality assume $0 < \sigma_{\mathrm{osc}} < \sigma_*
< \sigma_1$. Then in the hilltop region:
\begin{equation}
 \sigma_{\mathrm{osc}} 
 \gg \sigma_1 - \sigma_{\mathrm{osc}}
,  \qquad 
V_0  \gg m_\sigma^2 (\sigma_1 - \sigma_{\mathrm{osc}})^2,
\label{hilltop_limit}
\end{equation}
(the first condition corresponds to (\ref{Gu28}))
the power spectrum of linear perturbations becomes
\begin{equation}
 \mathcal{P}_\zeta \simeq
 \left( \frac{3 \hat{r}}{4 + 3 \hat{r}} \frac{\sigma_1 -
  \sigma_{\mathrm{osc}}}{\sigma_1 - \sigma_*}\frac{H_*}{2 \pi
  \sigma_{\mathrm{osc}}}  \right)^2,
\label{Pzetahilltop}
\end{equation}
where $\hat{r}$ is defined in (\ref{hatr}). 
The field values during inflation and at the onset of the oscillations are
related by
\begin{equation}
 \log \left(\frac{\sigma_1 - \sigma_*}{\sigma_1 -
      \sigma_{\mathrm{osc}}}\right)
 \simeq - A \frac{\sigma_{\mathrm{osc}}}{ \sigma_1 -
 \sigma_{\mathrm{osc}} },
\label{sigmas_hilltop}
\end{equation}
where $A$ is constant of order unity depending on whether reheating
happens before/after the curvaton starts to oscillate. The left hand side
of (\ref{sigmas_hilltop}) being a log term shows that
$\sigma_{\mathrm{osc}}$ is insensitive to $\sigma_*$, i.e.,
as one takes the hilltop limit $\sigma_* \to \sigma_1$, the value of 
$\sigma_{\mathrm{osc}}$ approaches $\sigma_1$ much slower than
$\sigma_*$ does. Hence $(\sigma_1 - \sigma_{\mathrm{osc}}) / (\sigma_1 -
\sigma_*)  \gg 1 $ and the power spectrum (\ref{Pzetahilltop}) is
enhanced compared to the quadratic case~(\ref{Pzetaquad}). 

However, we should also remark that the non-Gaussianity also mildly
increases in the hilltop limit as
\begin{equation}
 f_{\mathrm{NL}} \simeq \frac{5 (4 + 3 \hat{r})}{18 \hat{r}}
 \frac{\sigma_{\mathrm{osc}}}{\sigma_1 - \sigma_{\mathrm{osc}}} ,
\end{equation}
which is clearly much greater than unity from~(\ref{hilltop_limit}) even
when $\hat{r} \gg 1$. 
(For e.g., the typical value for $f_{\mathrm{NL}}$ from an axionic
curvaton at the hilltop is a few tens~\cite{Kawasaki:2011pd,Kawasaki:2012gg}.)

Thus we have seen that hilltop models that are free from the issues
discussed in Sec.~\ref{subsec:quad_curv} are now strictly
constrained from the latest $f_{\mathrm{NL}}$ bounds provided by Planck.
We also note that the increase of non-Gaussianity towards the hilltop is a
rather generic feature of hilltop spectator field models where the
curvature perturbations are generated after the field starts 
oscillating. (However, if the perturbations are produced before the
oscillations (as in most modulated reheating scenarios), then spectator
fields at the hilltop do not necessarily lead to large~$f_{\mathrm{NL}}$.)

Before concluding this section, let us remark that the $f_{\mathrm{NL}}$
constraints on hilltop curvatons can be resolved if the curvaton
initially lies extremely close to the potential maximum such that it 
dominates the Universe in the post-inflationary era before the
oscillations.\footnote{Such inflating curvatons can also be considered as a
variant of double inflation~\cite{Polarski:1992dq}.}
Such curvatons driving a (short) second inflationary epoch tend to 
generate small~$f_{\mathrm{NL}}$ (say, of
order unity) compatible with observations. 
Detailed discussions on inflating curvatons can be found in,
e.g. appendix of~\cite{Kawasaki:2012gg}.

\section{Discussion and conclusions}

We have discussed implications of recent Planck results for
spectator field models, such as the curvaton, modulated reheating and so on, which have been often 
referred as models generating  large (local-type) non-Gaussianity.
It should be noted that 
spectator field models of these kinds can also predict $f_{\rm NL} \sim \mathcal{O}(1)$ in wide range of  
their parameter space, thus, although Planck data now very severely constrains the non-linearity parameter $f_{\rm NL}$,
 they are not excluded only by the argument of non-Gaussianity. 
Importantly, the spectral index $n_s$ is also precisely measured by Planck and a red-tilted one is strongly 
favored.  Since spectator field models tend to give a scale-invariant 
power spectrum,  the consideration of the spectral index, in combination with $f_{\rm NL}$, 
gives significant implications for spectator field models. 

Based on the formula \eqref{eq:index} presented in Sec.~\ref{sec:2}, we discussed 
two possible scenarios to have a red-tilted spectral index consistent with Planck results:
(i) large variation of Hubble parameter during inflation  as in the large field inflation model 
(ii) a spectator field potential with a relatively large negative mass squared.  
The former one resorts to the inflation model, on the other hand, the latter one 
relies on the spectator field sector to realize a red-tilted spectrum.
Some possible scenarios for the latter case were discussed
in~Sec.~\ref{subsec:Nms}. 
Furthermore, we have given detailed discussions for the curvaton model
and argued that a hilltop curvaton  can be viable, giving a red-tilted spectral index and $f_{\rm NL} \sim \mathcal{O}(1)$,  
when the curvaton field initially lies extremely close to the potential maximum to 
give a second inflationary epoch. 
Other ways to reconcile the curvaton mechanism with observations were
proposed in Sec.~\ref{subsec:possible}.

Lastly we briefly comment on yet another possibility  having a successful spectator
field model. In fact, the constraint on the spectral index given in \eqref{ns} can be changed 
when we consider an extension of  the concordance cosmological model, the standard 
$\Lambda$CDM model. 
The effective number of neutrino species is usually set to be
 $N_{\rm eff} =3.046$. However, if we add extra relativistic degrees of freedom to vary $N_{\rm eff}$, 
  a scale-invariant power spectrum may give a equally good fit to the
observations. According to the Planck analysis on the one-parameter extensions, 
 the scale-invariant power spectrum is marginally allowed at 2$\sigma$ level \cite{Ade:2013lta}. 
From theoretical view point, there are a lot of candidates of dark
radiation \cite{Kobayashi:2011hp,Chun:2000jr}. In particular, the present authors
showed that dark radiation is naturally produced in the context of the
modulated reheating scenario \cite{Kobayashi:2011hp}. In the scenario,
the decay rate of the inflaton must depend on a modulus, which in turn
implies that the inflaton must couple to the modulus and hence a
significant amount of the modulus is produced through the inflaton
decay. 
We also note that an extension with freely varying $Y_p$ may also allow a
scale-invariant spectrum. A large value of the Helium abundance $Y_p$ can  
give a similar effect as in the case of the extra radiation discussed above. 
By varying $Y_p$ freely, the scale-invariant spectrum is marginally 
allowed at 2$\sigma$ level \cite{Ade:2013lta}.

\section*{Acknowledgment}
We thank Teruaki Suyama for helpful conversations. 
TK is also grateful to Dick Bond for valuable discussions.
FT would like to thank the YITP at Kyoto University for the hospitality
during the YITP workshop YITP-W-12-21 on ``LHC vs Beyond the Standard Model''.
This work was supported by the Grant-in-Aid for Scientific Research
(No.~23740195 [TT] and No.~21740187 [MY]), 
the Grant-in-Aid for Scientific Research on Innovative Areas
(No.~24111702 [FT], No.~21111006 [FT], No.~23104008 [FT], and No.~24111706 [MY]),  
Scientific Research (A) (No.~22244030 and No.~21244033) [FT], 
and JSPS Grant-in-Aid for Young Scientists (B) (No. 24740135) [FT]. 
This work was also supported by World Premier International Center
Initiative (WPI Program), MEXT, Japan [FT].


\begin{thebibliography}{100}

\bibitem{Enqvist:2001zp}
  S.~Mollerach,
  Phys.\ Rev.\ D {\bf 42}, 313 (1990);
  A.~D.~Linde and V.~F.~Mukhanov,
  Phys.\ Rev.\ D {\bf 56}, 535 (1997)
  [astro-ph/9610219];
K.~Enqvist and M.~S.~Sloth,
Nucl.\ Phys.\ B {\bf 626}, 395 (2002)
[arXiv:hep-ph/0109214];
D.~H.~Lyth and D.~Wands,
Phys.\ Lett.\ B {\bf 524}, 5 (2002)
[arXiv:hep-ph/0110002];
T.~Moroi and T.~Takahashi,
Phys.\ Lett.\ B {\bf 522}, 215 (2001)
[Erratum-ibid.\ B {\bf 539}, 303 (2002)]
[arXiv:hep-ph/0110096].

\bibitem{Dvali:2003em}
  G.~Dvali, A.~Gruzinov, M.~Zaldarriaga,
  Phys.\ Rev.\  {\bf D69}, 023505 (2004)
  [astro-ph/0303591];
  L.~Kofman,
  [astro-ph/0303614].

\bibitem{Ade:2013rta} 
  P.~A.~R.~Ade {\it et al.}  [ Planck Collaboration],
  arXiv:1303.5082 [astro-ph.CO].  

\bibitem{Ade:2013tta} 
  P.~A.~R.~Ade {\it et al.}  [ Planck Collaboration],
  arXiv:1303.5084 [astro-ph.CO].  

\bibitem{Lyth:2001nq} 
  D.~H.~Lyth and D.~Wands, in Ref.\cite{Enqvist:2001zp}

\bibitem{Kobayashi:2011hp} 
  T.~Kobayashi, F.~Takahashi, T.~Takahashi and M.~Yamaguchi,
  JCAP {\bf 1203}, 036 (2012)  [arXiv:1111.1336 [astro-ph.CO]].  

\bibitem{Hinshaw:2012fq} 
  G.~Hinshaw, D.~Larson, E.~Komatsu, D.~N.~Spergel, C.~L.~Bennett, J.~Dunkley, M.~R.~Nolta and M.~Halpern {\it et al.},
  arXiv:1212.5226 [astro-ph.CO].
  
\bibitem{Kawasaki:2011pd} 
  M.~Kawasaki, T.~Kobayashi and F.~Takahashi,
  Phys.\ Rev.\ D {\bf 84}, 123506 (2011)  [arXiv:1107.6011 [astro-ph.CO]].  

\bibitem{Kawasaki:2012gg} 
  M.~Kawasaki, T.~Kobayashi and F.~Takahashi,
  JCAP {\bf 1303}, 016 (2013)  [arXiv:1210.6595 [astro-ph.CO]].

\bibitem{Suyama:2010uj} 
  T.~Suyama, T.~Takahashi, M.~Yamaguchi and S.~Yokoyama,
 JCAP {\bf 1012}, 030 (2010)  [arXiv:1009.1979 [astro-ph.CO]].

\bibitem{Byrnes:2006fr} 
  C.~T.~Byrnes and D.~Wands,
  Phys.\ Rev.\ D {\bf 74}, 043529 (2006)
  [astro-ph/0605679].


\bibitem{Kawasaki:2008mc} 
  M.~Kawasaki, K.~Nakayama and F.~Takahashi,
  JCAP {\bf 0901}, 026 (2009)
  [arXiv:0810.1585 [hep-ph]];


\bibitem{Lyth:2002my}
  D.~H.~Lyth, C.~Ungarelli and D.~Wands,
  Phys.\ Rev.\  D {\bf 67}, 023503 (2003)
  [arXiv:astro-ph/0208055];

\bibitem{Lyth:2003ip}
  D.~H.~Lyth and D.~Wands,
  Phys.\ Rev.\  D {\bf 68}, 103516 (2003)
  [arXiv:astro-ph/0306500];

\bibitem{Bartolo:2003jx}
  N.~Bartolo, S.~Matarrese and A.~Riotto,
  Phys.\ Rev.\  D {\bf 69}, 043503 (2004)
  [arXiv:hep-ph/0309033];

\bibitem{Sasaki:2006kq} 
  M.~Sasaki, J.~Valiviita and D.~Wands,
  Phys.\ Rev.\ D {\bf 74}, 103003 (2006)  [astro-ph/0607627].  

\bibitem{Zaldarriaga:2003my}
  M.~Zaldarriaga,
  Phys.\ Rev.\  D {\bf 69}, 043508 (2004)
  [arXiv:astro-ph/0306006].

\bibitem{Suyama:2007bg}
  T.~Suyama and M.~Yamaguchi,
  Phys.\ Rev.\  D {\bf 77}, 023505 (2008)
  [arXiv:0709.2545 [astro-ph]].

\bibitem{Suyama:2013nva} 
  T.~Suyama, T.~Takahashi, M.~Yamaguchi and S.~Yokoyama,
  arXiv:1303.5374 [astro-ph.CO].  

\bibitem{Maldacena:2002vr} 
  J.~M.~Maldacena,
  JHEP {\bf 0305}, 013 (2003)  [astro-ph/0210603].  

\bibitem{Creminelli:2004yq} 
  P.~Creminelli and M.~Zaldarriaga,
  JCAP {\bf 0410}, 006 (2004)  [astro-ph/0407059].  

\bibitem{Bernardeau:2002jf} 
  F.~Bernardeau and J.~-P.~Uzan,
  Phys.\ Rev.\ D {\bf 67}, 121301 (2003)
  [astro-ph/0209330].

\bibitem{Bernardeau:2004zz} 
  F.~Bernardeau, L.~Kofman and J.~-P.~Uzan,
  Phys.\ Rev.\ D {\bf 70}, 083004 (2004)
  [astro-ph/0403315].

\bibitem{Lyth:2005qk} 
  D.~H.~Lyth,
  JCAP {\bf 0511}, 006 (2005)
  [astro-ph/0510443].

\bibitem{Salem:2005nd}
  M.~P.~Salem,
  Phys.\ Rev.\  D {\bf 72}, 123516 (2005)
  [arXiv:astro-ph/0511146].
  
\bibitem{Alabidi:2006wa}
  L.~Alabidi and D.~Lyth,
  JCAP {\bf 0608}, 006 (2006)
  [arXiv:astro-ph/0604569].

\bibitem{Kawasaki:2009hp} 
  M.~Kawasaki, T.~Takahashi, S.~Yokoyama and ,
  JCAP {\bf 0912}, 012 (2009)
  [arXiv:0910.3053 [hep-th]].

\bibitem{229378} 
  A.~A.~Starobinsky,
  JETP Lett.\ \ {\bf 42}, 152  (1985)
  [Pisma Zh.\ Eksp.\ Teor.\ Fiz.\ \ {\bf 42}, 124  (1985)].

\bibitem{astro-ph/9507001} 
  M.~Sasaki and E.~D.~Stewart,
  Prog.\ Theor.\ Phys.\ \ {\bf 95}, 71  (1996)
  [astro-ph/9507001].

\bibitem{astro-ph/0003278} 
  D.~Wands, K.~A.~Malik, D.~H.~Lyth and A.~R.~Liddle,
  Phys.\ Rev.\ D\ {\bf 62}, 043527  (2000)
  [astro-ph/0003278].

\bibitem{astro-ph/0411220} 
  D.~H.~Lyth, K.~A.~Malik and M.~Sasaki,
  JCAP\ {\bf 0505}, 004  (2005)
  [astro-ph/0411220].

\bibitem{Finelli:2010sh} 
  F.~Finelli, G.~Marozzi, A.~A.~Starobinsky, G.~P.~Vacca and G.~Venturi,
  Phys.\ Rev.\ D {\bf 82}, 064020 (2010)
  [arXiv:1003.1327 [hep-th]].

\bibitem{Sasaki:1995aw} 
  M.~Sasaki and E.~D.~Stewart,
  Prog.\ Theor.\ Phys.\  {\bf 95}, 71 (1996)
  [astro-ph/9507001].

\bibitem{Polarski:1994rz} 
  D.~Polarski and A.~A.~Starobinsky,
  Phys.\ Rev.\ D {\bf 50}, 6123 (1994)
  [astro-ph/9404061].

\bibitem{Kobayashi:2013nva} 
 T.~Kobayashi, R.~Kurematsu and F.~Takahashi,
 arXiv:1304.0922 [hep-ph].

\bibitem{Dterm} 
  K.~Kadota and M.~Yamaguchi,
  Phys.\ Rev.\ D {\bf 76}, 103522 (2007)  [arXiv:0706.2676 [hep-ph]];  
  K.~Kadota, T.~Kawano and M.~Yamaguchi,
  Phys.\ Rev.\ D {\bf 77}, 123516 (2008)  [arXiv:0802.0525 [hep-ph]].  

\bibitem{Fterm} 
  M.~Kawasaki, M.~Yamaguchi and T.~Yanagida,
  Phys.\ Rev.\ Lett.\  {\bf 85}, 3572 (2000)  [hep-ph/0004243]; 
  M.~Kawasaki, M.~Yamaguchi and T.~Yanagida,
  Phys.\ Rev.\ D {\bf 63}, 103514 (2001)  [hep-ph/0011104].  

\bibitem{mixed}
  T.~Moroi, T.~Takahashi and Y.~Toyoda,
  Phys.\ Rev.\  D {\bf 72}, 023502 (2005)
  [arXiv:hep-ph/0501007];
  T.~Moroi and T.~Takahashi,
  Phys.\ Rev.\  D {\bf 72}, 023505 (2005)
  [arXiv:astro-ph/0505339];
  K.~Ichikawa, T.~Suyama, T.~Takahashi and M.~Yamaguchi,
  Phys.\ Rev.\  D {\bf 78}, 023513 (2008)
  [arXiv:0802.4138 [astro-ph]];
K.~Ichikawa, T.~Suyama, T.~Takahashi and M.~Yamaguchi,
  Phys.\ Rev.\  D {\bf 78}, 063545 (2008)
  [arXiv:0807.3988 [astro-ph]].

\bibitem{Kobayashi:2013awa} 
  T.~Kobayashi and T.~Takahashi,
  arXiv:1303.0242 [astro-ph.CO].

\bibitem{ArmendarizPicon:1999rj} 
  C.~Armendariz-Picon, T.~Damour and V.~F.~Mukhanov,
  Phys.\ Lett.\ B {\bf 458}, 209 (1999)  [hep-th/9904075];  

\bibitem{Silverstein:2003hf} 
  E.~Silverstein and D.~Tong,
  Phys.\ Rev.\ D {\bf 70}, 103505 (2004)  [hep-th/0310221];  
  M.~Alishahiha, E.~Silverstein and D.~Tong,
  Phys.\ Rev.\ D {\bf 70}, 123505 (2004)  [hep-th/0404084].  
 
\bibitem{Baumann:2006cd} 
  D.~Baumann and L.~McAllister,
  Phys.\ Rev.\ D {\bf 75}, 123508 (2007)  [hep-th/0610285].  

\bibitem{Lidsey:2007gq} 
  J.~E.~Lidsey and I.~Huston,
  JCAP {\bf 0707}, 002 (2007)
  [arXiv:0705.0240 [hep-th]].

\bibitem{Kobayashi:2007hm} 
  T.~Kobayashi, S.~Mukohyama and S.~Kinoshita,
  JCAP {\bf 0801}, 028 (2008)
  [arXiv:0708.4285 [hep-th]].

\bibitem{Langlois:2009jp}
  D.~Langlois and L.~Sorbo,
  JCAP {\bf 0908}, 014 (2009)
  [arXiv:0906.1813 [astro-ph.CO]].

\bibitem{Nakayama:2011bc} 
  K.~Nakayama and T.~Suyama,
  Phys.\ Rev.\ D {\bf 84}, 063520 (2011)
  [arXiv:1107.3003 [astro-ph.CO]].

\bibitem{Kawasaki:2011zi} 
  M.~Kawasaki and T.~Takesako,
  Phys.\ Lett.\ B {\bf 711}, 173 (2012)
  [arXiv:1112.5823 [hep-ph]];
  Phys.\ Lett.\ B {\bf 718}, 522 (2012)
  [arXiv:1208.1323 [hep-ph]];
  M.~Kawasaki, F.~Takahashi and T.~Takesako,
  arXiv:1211.4921 [hep-ph].

\bibitem{Enqvist:2005pg} 
  K.~Enqvist and S.~Nurmi,
  JCAP {\bf 0510}, 013 (2005)
  [astro-ph/0508573].

\bibitem{Enqvist:2008gk} 
  K.~Enqvist and T.~Takahashi,
  JCAP {\bf 0809}, 012 (2008)
  [arXiv:0807.3069 [astro-ph]].

\bibitem{Enqvist:2009ww} 
  K.~Enqvist, S.~Nurmi, O.~Taanila and T.~Takahashi,
  JCAP {\bf 1004}, 009 (2010)
  [arXiv:0912.4657 [astro-ph.CO]].

\bibitem{Kobayashi:2012ba} 
  T.~Kobayashi and T.~Takahashi,
  JCAP {\bf 1206}, 004 (2012)
  [arXiv:1203.3011 [astro-ph.CO]].

\bibitem{Polarski:1992dq} 
  D.~Polarski and A.~A.~Starobinsky,
  Nucl.\ Phys.\ B {\bf 385}, 623 (1992).

\bibitem{Ade:2013lta} 
  P.~A.~R.~Ade {\it et al.}  [ Planck Collaboration],
  arXiv:1303.5076 [astro-ph.CO].

\bibitem{Chun:2000jr}
 E.~J.~Chun, D.~Comelli and D.~H.~Lyth,
 Phys.\ Rev.\ D {\bf 62}, 095013 (2000)
 [hep-ph/0008133];
K.~Ichikawa, M.~Kawasaki, K.~Nakayama, M.~Senami and F.~Takahashi,
JCAP {\bf 0705} (2007) 008
[arXiv:hep-ph/0703034];
K.~Nakayama, F.~Takahashi and T.~T.~Yanagida,
Phys.\ Lett.\ B {\bf 697} (2011) 275
[arXiv:1010.5693 [hep-ph]];
 W.~Fischler and J.~Meyers,
Phys.\ Rev.\ D {\bf 83}, 063520 (2011) 
[arXiv:1011.3501 [astro-ph.CO]];
 J.~Hasenkamp,
 Phys.\ Lett.\ B {\bf 707}, 121 (2012)
  [arXiv:1107.4319 [hep-ph]];
 J.~L.~Menestrina and R.~J.~Scherrer,
 Phys.\ Rev.\ D {\bf 85}, 047301 (2012)
 [arXiv:1111.0605 [astro-ph.CO]];
 D.~Hooper, F.~S.~Queiroz and N.~Y.~Gnedin,
 Phys.\ Rev.\ D {\bf 85}, 063513 (2012)
 [arXiv:1111.6599 [astro-ph.CO]];
 K.~S.~Jeong and F.~Takahashi,
 JHEP {\bf 1208}, 017 (2012)  [arXiv:1201.4816 [hep-ph]];
 M.~Cicoli, J.~P.~Conlon and F.~Quevedo,
arXiv:1208.3562 [hep-ph];
%
 T.~Higaki and F.~Takahashi,
JHEP {\bf 1211}, 125 (2012)  [arXiv:1208.3563 [hep-ph]];
 K.~Choi, K.~-Y.~Choi and C.~S.~Shin,
 Phys.\ Rev.\ D {\bf 86}, 083529 (2012)  [arXiv:1208.2496 [hep-ph]];
 P.~Graf and F.~D.~Steffen,
 arXiv:1208.2951 [hep-ph].  
 J.~Hasenkamp and J.~Kersten,
 arXiv:1212.4160 [hep-ph].  
 K.~J.~Bae, H.~Baer and A.~Lessa,
 arXiv:1301.7428 [hep-ph].  
 P.~Graf and F.~D.~Steffen,
 arXiv:1302.2143 [hep-ph];
 T.~Higaki, K.~S.~Jeong, F.~Takahashi and ,
 arXiv:1302.2516 [hep-ph].



\end{thebibliography}
\end{document}